# EFFECT OF DISORDER ON THE PEAK EFFECT AND STRUCTURE OF FLUX LINE LATTICE IN 2H-NbSe$_2$

S. S. Banerjee [1], N. G. Patil [1], Subir Saha [1], S. Ramakrishnan [1], A.K. Grover [1], S. Bhattacharya[1], G. Ravikumar [2], P.K. Mishra [2], T.V. Chandrasekhar Rao [2], V.C. Sahni [2], C.V. Tomy[3], G. Balakrishnan[3], D.Mck Paul [3] and M. J. Higgins [4]

[1]*Tata Institute of Fundamental Research, Mumbai - 400005, India*
[2]*TPPED, Bhabha Atomic Research Centre, Mumbai - 400085, India*
[3]*Department of Physics, University of Warwick, Coventry, CV4 7AL, U.K.*
[4]*NEC Research Institute, 4 Independence Way, Princeton, New Jersey 08540, USA*

The effect of disorder on flux line lattice ( FLL ) melting is studied via the Peak Effect phenomenon. On the upper branch of the Peak Effect curve, where this effect is more robust, we observe a stepwise disordering of the FLL, suggesting two first order transitions spanning the Peak Effect region. The lower, reentrant part of the curve is found to be strongly affected by disorder. Both the branch and the "nose", i.e., the turnaround of the curve, disappear in systems with stronger disorder and the observed results are consistent with the line separating a disentangled liquid or a glass from an entangled flux liquid.

PACS : 64.70 Dv, 74.60 Ge, 74.25 Dw, 74.60 Ec, 74.60 Jg

## I. INTRODUCTION

The effects of disorder on the phase boundaries in the flux line lattice (FLL) provide a way to measure the competition between inter-vortex interaction and both thermal and quenched disorder[1]. Due to the variety of materials with a large range of quenched disorder and transition temperatures, and the easy tunability of the interaction using the magnetic field, the FLL provides a rich testing ground of the theoretical concepts[2] and of the phase behaviour without and, more notably, with disorder[3]. In the case of the much-studied cuprates, quenched disorder is known to destroy[4] a sharp melting phase boundary of FLL; the residual effect is the presence of the so-called irreversibility line separating a less mobile state of FLL from a more mobile one for small driving forces. Whether the latter is indeed a phase transition line or a mere crossover of dynamic response remains uncertain[5]. A second question is concerned with the Peak Effect (PE), a sharp rise in the apparent critical current density in close proximity of the melting line. In some descriptions, this effect itself is associated with melting, while in some other, it represents a premelting softening of the FLL. The precise relationship between PE and FLL melting and whether it depends on specifics of systems under study continue to remain subjects of current work[6,7]. It was shown recently[8] that in the case of weak pinning single crystal samples of hexagonal 2H-NbSe$_2$, the PE curve bears a remarkable similarity to the theoretical predictions of Nelson[9] of the melting phase boundary of pure FLL, including the lower reentrant branch[2]. The location of this reentrant line, including the turnaround, often referred to as the "nose", is also in good quantitative agreement with simulations[10].

The disappearance of peak effect at field values below the nose was argued [8] to be consequence of dominance of pinning on the interaction. In order to further explore these ideas, we undertook investigations on the effects of increasing disorder on the magnetic signatures of the PE curve in 2H-NbSe$_2$. Specifically, we find that the entire low field part of this curve is extremely susceptible to disorder; neither the "nose" nor the lower branch of the PE curve survives stronger disorder. The upper branch is relatively more rugged; but even for this branch, over a wider parameter space, the full disordering of the FLL occurs via steps, suggestive of first order transitions and ending, remarkably, with the exact location for the disappearance of thermomagnetic history effect in the magnetization screening response.

## II. EXPERIMENTAL RESULTS

The experiments were carried out on a single crystal of Fe-doped $NbSe_2$, similar to samples where transport studies of metastability were reported recently[12]. Fig.1(a) shows a dc magnetization hysterisis loop for $H\|c$ at 5.1 K recorded with a standard Quantum Design Inc. Squid magnetometer. Note that the magnetization loop opens up anomalously at the onset of the PE at a field marked $H_m$, reaching a peak at $H_p$. The resulting ``magnetization bubble'' closes at a higher field $H_{irr}$ below the nominal $H_{c2}$. Fig.1(b) shows the H-dependence of the real part of the ac susceptibility $\chi'$. In this experiment, the PE is manifested as an anomalous enhancement of the diamagnetic response that appears at $H_m$ and reaches a maximum value at $H_p$. Together, these results (Figs 1(a) and 1(b)) show the self-consistency between the anomalous dc magnetization hysterisis and the ac susceptibility behaviour in detecting the PE.

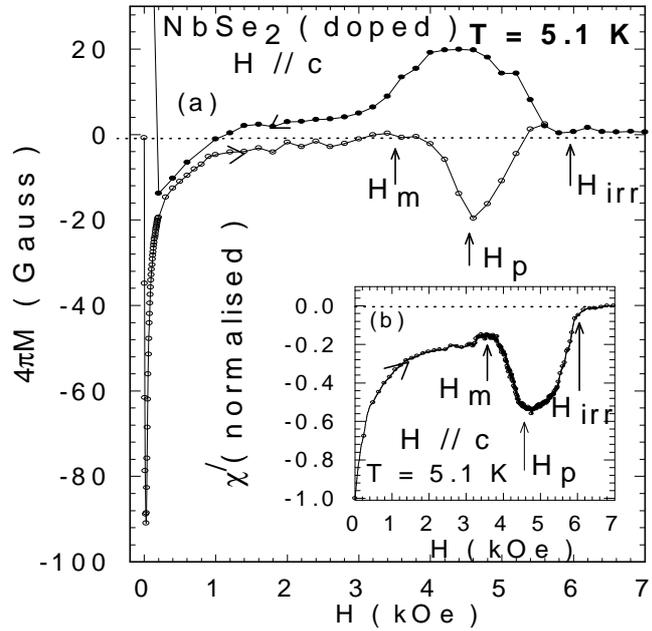

Fig.2 shows the temperature dependence of the real part of the ac susceptibility $\chi'$ at a fixed H in the doped sample for two different thermomagnetic histories : field cooled (FC) and zero-field-cooled (ZFC). For ZFC, we observe a sharp drop in $\chi'$ at the onset of the PE, marked $T_m$, following which $\chi'$ continues to decrease until at a higher temperature marked $T_p$ it shows yet another sharp drop, after which it rises rapidly. A comparison with the field cooled (FC) data shows a distinct difference with the ZFC data. The $\chi'$ response in FC state at low temperatures ($<T_m$) is more diamagnetic, implying a more strongly pinned (i.e., more disordered) lattice, in agreement with recent transport measurements[12]. Importantly, the thermomagnetic memory effect disappears at $T_p$ above which the FC and ZFC curves merge into one. The peak effect extends over a region spanned by two temperatures, $T_m$ and $T_p$. For ZFC, the more weakly pinned and hence the more ordered state, we see significant structure in $\chi'$, namely, the sudden drops, while the more disordered FC state shows much less structure. Since thermomagnetic history effect disappears above $T_p$, the lattice is disordered in equilibrium and hence a candidate for a pinned liquid or an amorphous FLL. Field cooling through this regime freezes FLL in a disordered structure of the vortices and thus the FC state is more akin to a supercooled liquid or a glass. Transport data[12] showed that the FC state, metastable in the peak regime and below, when "annealed" with a greater than critical current passed through it, produces a state with a critical current nearly identical to the ZFC state, i.e., the ZFC state is closer to the equilibrium structure of the system. Thus, the structures observed in the ZFC state closely approximate the equilibrium features of the system. In what follows, we present a tentative explanation of the observations in Fig. 2. First we note that both the drop features are extremely sharp and narrower than 5mK, whereas the normal to superconducting transition itself (see zero field curve in Fig.3(b)) is nearly fifty times as large. The two sudden drop features are thus signatures of well-defined first order phase transitions. The sudden drop in $\chi'$ for ZFC at $T_m$ is analogous to the sharp increase of $J_c$ at $H_m$ in Fig. 3 of Ref[12].

**Figure 1:** The main panel (a) shows dc magnetisation hysterisis curve for H//c in doped crystal of 2H-NbSe$_2$ ($T_c(0) \approx 6$ K) at 5.1 K. The inset (b) depicts the in-phase ac [f=211 Hz, $h_{ac}$=0.5 Oe (r.m.s)] susceptibility ($\chi'$) vs H at nearly the same temperature. The fields $H_p$ and $H_{irr}$ in the inset are marked as per dc data in the main panel.

Within the Larkin-Ovchinnikov[14] model of collective pinning, the pinning force $F_p$ is given by $F_p = (n_p f^2/V_c)^{1/2}$, where f is the elementary pinning interaction, $n_p$ is the density of pins and $V_c$ is the correlation volume ($= R_c^2 \cdot L_c$, where $R_c$ and $L_c$ are collective pinning radial and longitudinal lengths[2,3,14]). Since the pinning interaction can only decrease with increasing T, a sharp increase in $F_p$ at $T_m$ can only occur via the sudden drop in $V_c$ and it must come from a sudden drop in one (or more) of the elastic moduli of FLL. At a higher T, we observe a second drop in $V_c$ at $T_p$, which we attribute to the final disordering of the FLL through the collapse of another elastic constant. A plausible implication is that the two elastic constants are the shear ($c_{66}$) and tilt ($c_{44}$) moduli. One can envisage two different scenarios : the first, proposed by Kes and coworkers[15], suggests that the first drop in $V_c$ corresponds to the collapse of the tilt modulus and the crossover of the effective dimensionality of the FLL from two to three. Since it was not observed before, the second anomalous drop in $V_c$ is not explained in this case. A second scenario would suggest that the onset (near $T_m$) of the PE is caused by the collapse of $c_{66}$; above $T_m$, one has a line liquid which is still correlated along H. In that case, the second anomaly at $T_p$ could be the collapse of the tilt modulus $c_{44}$, which completes the disordering of the FLL. Above $T_p$, one has a fully disordered state which could explain the disappearance of the thermomagnetic history effect above $T_p$. Which (if any) of the scenarios is correct is unclear at present, but it is nevertheless obvious that the full disordering of the FLL does not happen in one step.

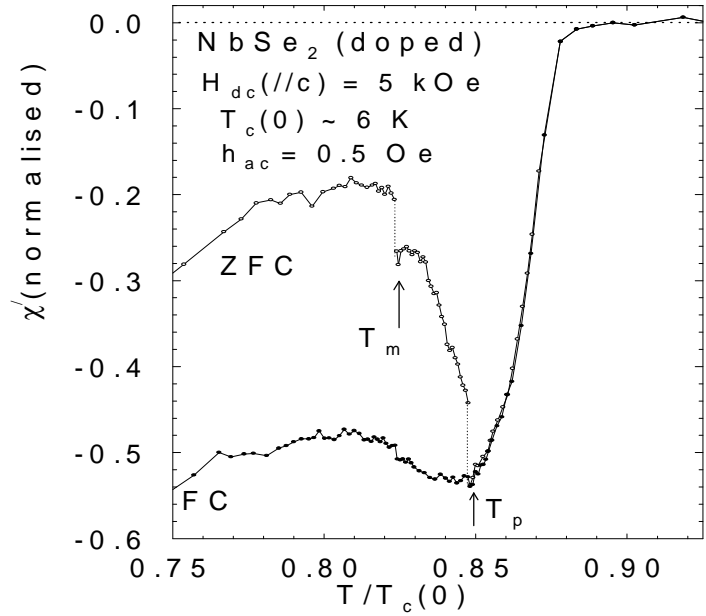

**Figure 2 :** Temperature variation of the in-phase ac susceptibility $\chi'$ at H(//c)=5 kOe in the doped crystal of 2H-NbSe$_2$ for two thermomagnetic histories, zero field cooled (ZFC) and field cooled (FC). The arrow mark the temperatures $T_m$ and $T_p$ at which discontinuous jumps are observed in ZFC case. The $\chi'$ data are normalised such that the extremum (diamagnetic) screening response corresponds to the value 1.

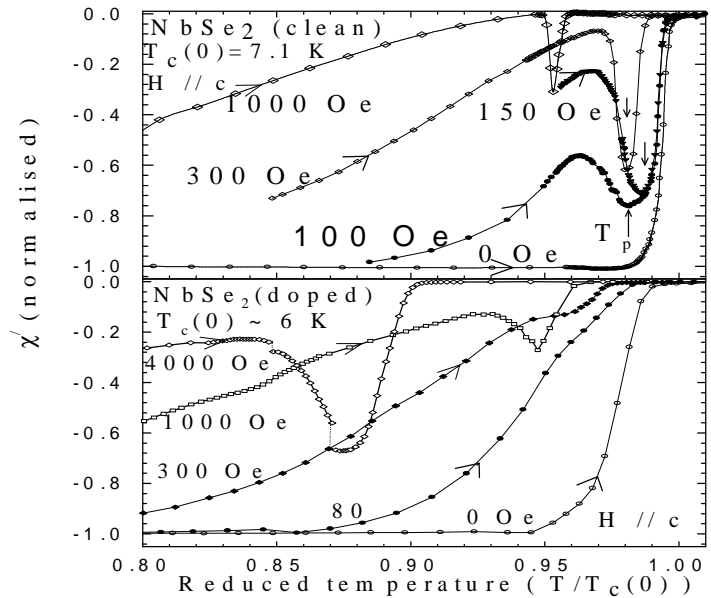

**Figure 3:** Temperature variation of $\chi'$ at different H (//c and ZFC) in the low field region for a clean and a doped crystal of 2H-NbSe$_2$. The doped crystal is more disordered than the undoped clean specimen as it has lower value of $T_c(0)$ and a broader transition in zero field. The vertical arrows in Fig. 3(a) mark the peak temperatures $T_p$(H).

Figs.3(a) and 3(b) show isofield measurements of $\chi'$ in an earlier[8,16] clean crystal specimen and the present doped crystal, respectively. For the cleaner system ( in Fig. 3(a) ), the data span fields greater than and less than the "nose" location ( cf. Fig. 4(b) ); one observes a sharper peak effect and the nonmonotonic variation of $T_p$ (marked by arrows) with H leading to the reentrance characteristic in PE curve as shown in Fig. 4(b). In the doped system ( see Fig. 3(b) ), however, the peak effect starts to broaden at significantly higher fields (than the "nose" location), it manifests as a shoulder in $\chi'$ around 150 gauss and for further lower values, it can no longer be located precisely. It is important to note that the gradual broadening of the PE transition and its ultimate disappearance are precisely what would be expected for the increasing effects of disorder in destroying the phase boundary defining the melting of FLL[5,17]. Also, there is no nonmonotonicity in the field range where the PE is observed in the doped sample. In other words, neither the lower branch, nor the nose can be detected unambiguously for the more disordered system.

**III. DISCUSSION**

The (H,T) phase boundaries emerging from ac and dc magnetization experiments for both doped and clean samples are shown in Fig. 4. The two PE lines in Fig. 4(a) for doped sample correspond to the two jumps in $\chi'$ at $T_m$ and $T_p$ at fixed H (cf. Fig.2). Clearly, a simple Lindemann criterion of a one step melting process cannot be applied to the two lines. Nevertheless, we attempt a fit to the centre of the two lines, for lower T and higher H, i.e., the so called upper branch of the PE curve, for an order of magnitude estimate of Lindemann parameter ($c_l$) by using the relation 5.5 of Ref. 2, $B_m(T) = \beta_m (c_l^4 / Gi) H_{c2}(0)(1 - T/T_c)^2$. A fit to this expression with values of various parameters kept the same as those in Ref. 8, yields a reasonable value for the Lindemann parameter ($c_l \sim 0.1$).

The PE becomes immeasurably weak at around 150 gauss for the doped case ( see Fig. 3(b) ), while in the clean system it disappears below the nose around 30 gauss. We note ( Fig. 4(b) ) that the upper branch of the PE curve is also affected by disorder: the curve shifts down with increasing disorder ( cf. $\chi'$ curves at H = 300 Oe in clean and doped samples in Fig. 3(a) and Fig. 3(b), respectively). This can be understood on general grounds since increased disorder reduces the correlation length of the vortex solid which can thus melt with weaker thermal fluctuations. The lower branch of PE curve is more profoundly altered. As stated earlier neither the nose, nor the lower branch is detectable in the more disordered doped case, which implies that the solid phase is squeezed upwards to higher H. In other words, larger intervortex interaction is needed to stabilise the solid phase. The disappearance of the PE curve at fields lower than 30 Oe even for the clean case was attributed [8] to the dominance of the disorder for the dilute array of vortices, as per arguments given for the Nelson-Le Doussal line[2,17]. The relevant boundary is governed by the entanglement length $L_E$ becoming equal to the pinning length $L_c$, following the relationship 6.47 of Ref. 2 :

$$\frac{L_E}{L_C} = \frac{\pi \kappa^2 \ln \kappa}{\sqrt{2}} \left[\frac{a_o}{2\pi\kappa(0)}\right]^2 \left[\frac{j_c}{G_i j_o}\right]^{1/2} \frac{(1-t)^{4/3}}{t}$$

where $a_0^2 \propto 1/B$ and the rest of the symbols have their usual meaning[2]. In the clean sample, it was found that $L_c / L_E \approx 1$ at $B \approx 30$ Oe[8]. The relation (1) implies that the crossover field (corresponding to $L_c / L_E = 1$) is proportional to the square root of the critical current density $j_c$, which is larger for the doped case by nearly a factor of 50. This coupled with the nominal changes in the values of $\kappa$ and $\lambda(0)$ of the doped sample as compared to those in a clean sample yields a crossover field of $\sim 4\times10^2$ Oe at $t \approx 0.96$, in reasonable agreement with the observed value of 150 Oe (cf. Fig 3(b)). Thus, the loss of the PE curve is consistent with the transformation of a thermally entangled flux liquid at low fields to a pinning dominated disentangled liquid or a glass[18]. Whether this crossover is accompanied by changes in the dynamic response of the system remains unknown.

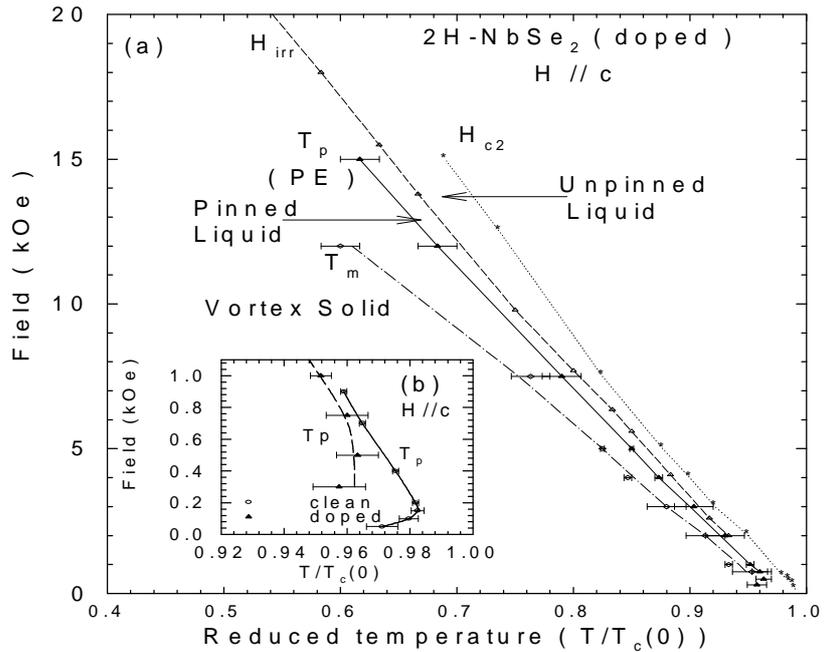

**Figure 4 :** Magnetic phase diagram for H//c in doped crystal of $2H\text{-}NbSe_2$ having $T_c(0) \sim 6$ K. The two PE lines, $T_p(H)$ and $T_m(H)$, and the irreversiblity line $H_{irr}(T)$ can be identified from the data such as in Fig. 2 and Fig.1, respectively. The inset (b) shows a comparision between $T_p(H)$ lines in the doped and a clean crystal. The $T_p(H)$ line in the latter sample displays the nose and the reentrant characteristics at $H < 150$ Oe.

The irreversibility line ($H_{irr}$) in Fig.4(a), has been obtained from data such as in Fig.1. Below this line (and above $T_p$ line), the FLL is disordered but pinned and is thus very likely a candidate for pinned liquid state as mentioned above. The irreversibility line thus corresponds to a crossover between a pinned and an unpinned liquid, as the correlation time of a very viscous flux liquid (plastic time $t_{pl}$ as mentioned in [2]) becomes comparable to the pinning time $t_{pin}$.

## IV. CONCLUSION

In conclusion, we have observed a stepwise disordering of the FLL in the upper branch of the peak effect curve in a more disordered sample of $2H\text{-}NbSe_2$. In a cleaner sample these steps, if present, are presumably too close to be resolved. We also find that the role of disorder is to shrink the solid phase[19] in the H-T space for both branches . For the lower reentrant branch in particular, the effect of disorder is very dramatic in the form of a disappearance of the "nose" and the entire branch, due, presumably, to the intersection of the Nelson Le Doussal line with the melting line of FLL at higher field values[2,17], below which the entangled flux liquid phase is destroyed by disorder.